# Low temperature properties of a new Kondo Lattice compound $Yb_2Ir_3Sn_5$

Y. Singh, R. S. Sannabhadti and S. Ramakrishnan

Tata Institute for Fundamental Research,
Homi Bhabha Road, Mumbai, 400 005

## Abstract

*We report the preparation and results of our magnetic and transport measurements on a new Yb-based ternary compound $Yb_2Ir_3Sn_5$. This compound forms in the $Cmc2_1$ structure. DC magnetic susceptibility between 1.8 K and 300 K reveal that Yb is in its magnetic trivalent state in this compound. Electrical resistivity measurements down to 1.5K show a behavior typical of Kondo lattice compounds with a minimum in resistivity around 8K and a coherence peak near 3K. AC magnetic susceptibility down to 100mK reveals an anti-ferromagnetic transition below 1.35 K.*

## INTRODUCTION

We have earlier shown that compounds of the type $R_2T_3X_5$, where R=rare-earth, T=transition metal and X=Si or Ge, form in various structures and show interesting superconducting and magnetic properties at low temperatures [1-5]. Ce and Yb based compounds in particular are expected to show interesting low temperature behavior which basically arise due to the competition between the Kondo effect, which tries to quench the rare-earth magnetic moments, and between the RKKY interaction which favors long range magnetic order. Depending on the relative strengths of these two interactions various ground states like heavy fermion behavior, Kondo effect without magnetic ordering or magnetic ordering of Kondo reduced moments can be encountered.

Whereas a lot of investigations have been done on various Ce based compounds [6-8], a similar effort is lacking for Yb based compounds due to the difficulty encountered in the preparation of Yb based compounds because of the high vapor pressure of Yb.

We have recently turned our efforts into trying to prepare good quality samples of Yb based compounds of the type $Yb_2T_3X_5$ (where T is a transition metal and X is an s, p element like Si or Ge) and search for compounds with interesting low temperature properties [1, 2]. Here we report the preparation and preliminary electrical transport and magnetic property measurements on a new Yb based ternary compound $Yb_2Ir_3Sn_5$.

## EXPERIMENTAL

Polycrystalline samples of $Yb_2Ir_3Sn_5$ were prepared by arc-melting on a water cooled copper hearth under argon atmosphere. First the binary $Ir_3Sn_5$ was prepared by arc-melting. This binary was then crushed into a fine powder and small pieces of Yb were mixed with this powder. The whole assembly was then pressed into a closely packed pellet. This pellet was then arc-melted several times to get the final $Yb_2Ir_3Sn_5$ compound. Powder X-ray diffraction on the as cast samples were performed. The diffraction pattern could be indexed (except a few small intensity peaks which indicate less than 3% of a second phase) to the orthorhombic $Cmc2_1$ structure with lattice constants a=4.365Å, b=26.132Å and c=7.158Å.

The AC magnetic susceptibility between 100mK and 4K and the DC magnetic susceptibility between 1.8K and 300K were measured using an adiabatic demagnetization system (CMR, Cambridge) and a commercial SQUID magnetometer respectively. The electrical resistivity between 1.5K and 300K was measured on a home built setup using an LR-700 AC resistance bridge. The standard 4-probe method was employed with contacts made using silver paste.

## RESULTS

**Fig.1** shows the temperature dependence of the AC and DC magnetic susceptibility for $Yb_2Ir_3Sn_5$. The main panel of the figure shows $1/\chi(T)$ data between 1.8K and 300K. The solid line is a fit to a modified Curie-Weiss expression given by $\chi = \chi_0 + C/(T-\theta)$, where $\chi_0$ is a temperature independent term, C is the Curie constant and $\theta$ is the Curie-Weiss temperature. From the fit we obtain $\theta$=-42 K and C= 5.49. From this value of C the effective moment of Yb can be estimated and comes out to be $\mu_{eff} = 4.56\mu_B$ which is very close to the value of

4.54 $\mu_B$ expected for trivalent Yb ions. This indicates that Yb is in its $3^+$ and hence magnetic valence state in this compound. A large and negative value of θ indicates the presence of strong hybridization between the localized Yb moments and the conduction electron spins. No signature of magnetic ordering is found down to 1.8 K. However, AC susceptibility (χ) measurements down to 100mK reveal that the compound undergoes an anti-ferromagnetic transition below 1.35 K (inset of **Fig.1**).

Thus, our magnetic measurements reveal that $Yb_2Ir_3Sn_5$ undergoes anti-ferromagnetic ordering of $Yb^{3+}$ moments below 1.35 K and that hybridization effects are indicated from the large and negative value of the Curie-Weiss temperature θ.

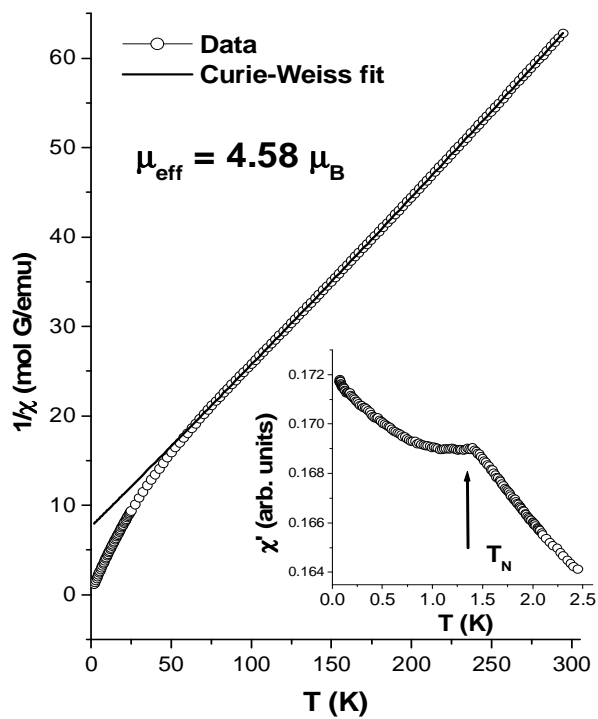

**Fig 1** Temperature dependence of the AC (see inset) and DC magnetic susceptibility between 100mK and 300 K.

**Fig.2** shows the electrical resistivity of $Yb_2Ir_3Sn_5$ between 1.5 K and 50 K to highlight the low temperature behavior. It can be seen that the resistance monotonically decreases till it reaches a shallow minimum around 8 K after which there is an increase in the resistance with decreasing temperature. A maximum is reached at 3 K after which the resistance falls rapidly with decreasing temperature. These features are hallmarks of concentrated Kondo lattice systems where the upturn in the resistance come about due to the scattering of the conduction electrons from the $Yb^{3+}$ magnetic moments and the maximum in resistance is seen at the onset of coherent scattering of the conduction electrons from Kondo centers arranged periodically on the lattice.

Thus, our resistivity measurements show that $Yb_2Ir_3Sn_5$ behaves like a typical Kondo lattice system below 10 K.

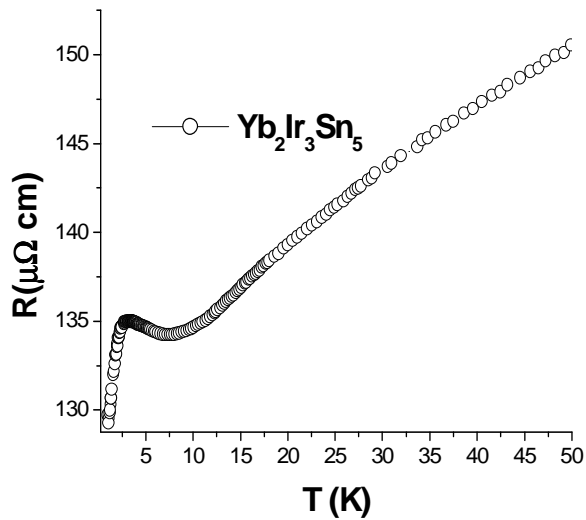

**Fig 2** The resistivity of $Yb_2Ir_3Sn_5$ between .4 K and 50 K (see text for details).

## CONCLUSION

We have studied the low temperature properties of a new Yb based ternary compound $Yb_2Ir_3Sn_5$. Our low temperature magnetic susceptibility and electrical resistivity measurements reveal that $Yb^{3+}$ moments undergo anti-ferromagnetic ordering transition below 1.35K and below 10K it behaves like a typical Kondo lattice system with probably a low Kondo temperature $T_K$=5 K.

It would be interesting to perform heat capacity measurements down to 0.3 K to look at the bulk nature of the magnetic transition and also to get an estimate of the Sommerfeld's coefficient γ to see if heavy fermion behavior is observed.

Also, AC susceptibility measurements in a DC magnetic field have indicated a reduction in the magnetic transition temperature. Therefore, it should in principal be possible to apply large enough magnetic fields and see if we can push the magnetic transition continuously to zero temperature and reach a "field induced quantum critical point".

These measurements are in progress and will be reported elsewhere.